\documentclass[12pt]{article}
\usepackage[dvips]{graphicx}
\usepackage{latexsym}
\def\Comment#1{}
\newcommand{\bean}{\begin{eqnarray*}}
\newcommand{\eean}{\end{eqnarray*}}

\newcommand{\gapproxeq}{\lower
.7ex\hbox{$\;\stackrel{\textstyle >}{\sim}\;$}}
\newcommand{\lapproxeq}{\lower
.7ex\hbox{$\;\stackrel{\textstyle <}{\sim}\;$}}

\newcommand\lsim{\mathrel{\rlap{\lower4pt\hbox{\hskip1pt$\sim$}}
    \raise1pt\hbox{$<$}}}
\newcommand\gsim{\mathrel{\rlap{\lower4pt\hbox{\hskip1pt$\sim$}}
    \raise1pt\hbox{$>$}}}
\newcommand{\ba}{\begin{array}}
\newcommand{\ea}{\end{array}}

\newcommand{\be}{\begin{equation}}
\newcommand{\ee}{\end{equation}}
\newcommand{\bear}{\begin{eqnarray}}
\newcommand{\eear}{\end{eqnarray}}
\newcommand{\tab}{\hspace*{0.5cm}}

\newcommand{\ket}{\,\rangle}
\newcommand{\bra}{\langle \,}
\newcommand{\eqn}[1]{(\ref{#1})}
\newcommand{\cO}{{\cal O}}
\newcommand{\bel}[1]{\be\label{#1}}
\newcommand{\mL}{\mathcal{L}}

\newcommand{\mM}{\mathcal{M}}

\newcommand{\Frac}[2]{\frac{\displaystyle #1}{\displaystyle #2}}

\begin{document}
\thispagestyle{empty}
\begin{titlepage}
\begin{center}
\hspace*{10cm} {\bf IFIC/04-07   } \\
\hspace*{10cm} {\bf FTUV/04-0805   }\\
\vspace*{2.75cm} 
\begin{Large}
{\bf Pion and Kaon Decay Constants: \\
Lattice vs. Resonance Chiral Theory  
}
 \\[2.4cm]
\end{Large}
{ \sc J.J. Sanz-Cillero } \\[0.8cm]

{\it Departament de F\'\i sica Te\`orica, IFIC, Universitat de Val\`encia -
CSIC\\
 Apt. Correus 22085, E-46071 Val\`encia, Spain }\\[0.5cm]

\today  
\vspace*{2cm}
\begin{abstract}
\noindent
The Lattice results for the pion and kaon decay constants
are analysed within the Resonance Chiral Theory framework in the large $N_C$
limit. The approximately 
linear behaviour  of the observable at large light-quark mass is explained  
through the interaction with the lightest multiplet of scalar resonances.  
The analysis of the Lattice results  
allows to obtain the resonance mass $M_S=1049\pm 25$ MeV and the Chiral
Perturbation Theory parameters at leading order in $1/N_C$. 
\end{abstract}
\end{center}
\vfill

\eject
\end{titlepage}

\pagenumbering{arabic}

\parskip12pt plus 1pt minus 1pt
\topsep0pt plus 1pt
\setcounter{totalnumber}{12}

\section{Introduction}
\tab 
Quantum Chromodynamics (QCD) has been proved to be the proper theory 
to describe the strong interactions. However in the low energy region 
the theory in terms of quarks and gluons becomes highly non perturbative. 
These degrees of freedom  get confined within complex hadronic states.  
Below the first resonance multiplet ($E\ll M_\rho$),   the
only particles in the spectrum are the light octet of pseudo-scalars,
the pseudo-Nambu-Goldstone bosons (pNGB) 
from the spontaneous chiral symmetry breaking. 
At low momenta and small pNGB masses 
one may describe their interactions through a chiral invariant 
effective field theory, Chiral Perturbation Theory 
($\chi$PT)~\cite{chpt}. It establishes an expansion on 
powers of the external momenta and masses  over a characteristic chiral scale 
$\Lambda_\chi\sim 4 \pi F\sim 1.2$ GeV, being $F\simeq F_\pi=92.4$ MeV the 
physical pion decay constant. 

The chiral expansion breaks down when either the momenta or  the 
pNGB masses become large; when they approach to the $\rho(770)$ mass 
the chiral expansion 
fails.  
The mesonic resonances can then be  produced   
and their effects cannot be neglected 
any longer. 

Alternatively, it is possible to employ 
the $1/N_C$ expansion to describe the matrix
elements --being $N_C$ the number of colours--~\cite{NC}. 
Resonance Chiral Theory (R$\chi$T)~\cite{the role} 
incorporates the interactions 
between the resonances and the Goldstones at leading order in $1/N_C$
and  also implements the chiral  symmetry of the interaction.    
Likewise, $\chi$PT is fully  recovered when going to low
energies.

At leading order in $1/N_C$ (LO) the observables are given by the tree-level
amplitudes, being the mesonic loops suppressed by $1/N_C$~\cite{NC}. 
In that situation  R$\chi$T is able to reproduce the short 
distance behaviour 
required by QCD for the pion form factors, two-point 
Green functions and  forward scattering amplitudes~\cite{spin1fields,PI:02}. 
The $1/N_C$ counting can also be carried  to the next order (NLO) 
in a systematic way and the quantum loop corrections 
might be calculated~\cite{quantumloops}.

The present study is focused on   how 
the variation of the quark masses affects 
the pion and kaon decay constants, $F_\pi$ and $F_K$, under the R$\chi$T
framework and the the $1/N_C$ expansion.   
Lattice calculations 
have provided information about QCD results for unphysical values of the 
$u/d$ light quarks~\cite{latticedata}. 
The simulations are forced sometimes to work with masses of nearly the size of
the physical strange quark mass or higher. Thus, 
the usual $\chi$PT extrapolations break down 
and generate large unphysical chiral logarithms, yielding an 
large bending  in the extrapolation~\cite{latticedata,fpiextra,donoghue}. 
Nonetheless, 
more than purely numerical values, 
this work aims providing a possible way to analyse the   
Lattice simulations at  large quark masses,  explaining     
why the usual linear extrapolations  work so well and what 
are the underlying physical foundations.

Through the inclusion of the first resonance multiplets, with masses 
$M_R\sim 1$ GeV,    
one expects to reproduce the  physics for 
the Goldstones up to that range of momenta and masses. 
Moreover, at LO in $1/N_C$ only the scalar
resonances contribute to $F_\pi$ and $F_K$. Since the first multiplet 
of pseudo-scalar resonances is at 
much higher masses, it will not be considered in the calculation 
and its mixing with  the pNGB will be neglected.  

\section{R$\chi$T lagrangian at LO in $1/N_C$}
\tab
The  leading order resonance  lagrangian was developed  
in Ref.~\cite{the role} and  includes one multiplets of 
vector, axial-vector, scalar and pseudo-scalar
resonances. It is composed by a pair of terms including only Goldstones 
--the $\cO(p^2)$ $\chi$PT Lagrangian~\cite{chpt}--,
\bel{eq.L2}
\mL_2 \, = \, \Frac{F^2}{4} \bra u_\mu u^\mu 
\, + \, \chi_+ 
\ket \, \, ,
\ee
and pieces including as well resonance fields. The brackets $\bra ... \ket$
mean trace of the flavour matrices. The chiral tensors including the Goldstone
fields can be obtained in Refs.~\cite{the role,spin1fields,PI:02} 
and they contain the external
fields $v^\mu, \, a^\mu    , \, s$ and $p$, 
and the non-linear realization of
the chiral symmetry given by the tensor 
$u=\exp{\left(i \Phi /\sqrt{2} F\right)}$,  with the pNGB fields 
$\Phi=\sum_a  \phi_a \lambda_a /\sqrt{2} $.

In the large $N_C$ limit 
the $q\bar{q}$ resonances form $U(3)$ multiplets. 
The fields of a multiplet can be put together in a $3\times 3$ 
matrix which transforms linearly under the chiral symmetry,
\be
R \, = \, \sum_a \, R_a \, \Frac{\lambda_a}{\sqrt{2}}  \, ,
\ee
containing  one chiral singlet field $R_0$ and  the remaining octet 
fields $R_a$, with $a=1,...8$. All the fields  $R_a$  of the multiplet 
would be degenerate in the large $N_C$  and massless quark limit. 
The kinetic terms are then constructed with these matrices:
\bel{eq.LRkin} 
\ba{ccl}
\mL_R^{\mathrm{Kin}}(R = V,A) & = & 
    - {1\over 2}\, \langle \nabla^\lambda R_{\lambda\mu}                      
\nabla_\nu R^{\nu\mu} -{1\over 2} \, M^2_R R_{\mu\nu} R^{\mu\nu}\rangle
\,  ,
\\ \\
\mL_R^{\mathrm{Kin}}(R = S,P) & = & {1\over 2} \,
\langle \nabla^\mu R 
\nabla_\mu R - M^2_R R^2\rangle\, , 
\ea
\ee
and the interaction lagrangian, linear in the resonance
fields~\cite{the role},   
\bel{eq.LRint}
\ba{ccl}
\mL_{2V} & = &  \Frac{F_V}{ 2\sqrt{2}} \,
     \langle V_{\mu\nu} f_+^{\mu\nu}\rangle +                                   
    \Frac{i G_V}{ 2 \sqrt{2}} \, \langle V_{\mu\nu} 
\left[ u^\mu ,  u^\nu \right] \rangle ,                  
\\ 
\\
\mL_{2A} & = & \Frac{F_A}{ 2\sqrt{2}} \,                                                                    
    \langle A_{\mu\nu} f_-^{\mu\nu} \rangle  ,                                
\\
\\  
\mL_{2S}  & = &  c_d \, \langle S u_\mu            
u^\mu\rangle + c_m \, \langle S \chi_+ \rangle \,  ,                                
\\
\\    
\mL_{2P} & = &  id_m \, \langle P \chi_-                                                                          
\rangle \,  ,                           
\ea
\ee
where the vector and axial-vector resonances are described by  
antisymmetric tensors. 

\section{Scalar tadpole and field redefinition}
\tab
Analysing the LO lagrangian one observes the presence of a term linear 
in the scalar fields, i.e. a scalar tadpole.  
It is given by the term of $\mL_{2S}$ in Eq.~\eqn{eq.LRint} 
with the coupling  $c_m$, 
which also provides the vertex  for the scalar resonance production 
from a scalar quark current.

Chiral symmetry requires the quark masses to enter in the 
effective lagrangian  
only through the  tensor $\chi \, = \, 2 \, B_0 \, \{s(x)+ip(x)\}$, 
which  appears in the chiral covariant combinations 
$\chi_\pm= u^\dagger \chi u^\dagger \pm u \chi^\dagger u$. 
In order to recover physical QCD, 
the external fields are evaluated at the end of the calculation  
as $\chi=2 B_0 \mM$, being 
$\mM=$diag$(m_u,m_d,m_s)$ the diagonal matrix with the light quark masses.

The pieces of the lagrangian containing the scalar fields are
\be
\ba{rl}
\mL_R^{\mathrm{Kin}}(S)\, + \, \mL_{2S} \, \, =& 
\, \, \Frac{1}{2}\bra \partial^\mu S \partial_\mu S\ket 
\, - \, \Frac{1}{2} \, M_S^2 \, \bra\, S^2 \,\ket \,+ \, 
4 B_0 c_m \, \bra \, S \, \mM \, \ket \, + \, \cO(S^2\Phi^2,\, S\Phi^2) \, .
\ea
\ee
The scalar field has therefore a non-zero vacuum expectation 
value (v.e.v.).
In order to define the quantum field theory around 
the minimum one needs to perform in the scalar field the shift:
\be
S \, = \, \bar{S} \, + \,\Frac{ 4 B_0 c_m}{ M_S^{2}}\, \mM  \, , 
\ee
where the shifted scalar nonet fields $\bar{S}$ has a zero v.e.v.  
Nonetheless this shift makes $\bar{S}$ not to be chiral covariant any longer. 
For the present  work this
detail is not relevant although other alternative shifts 
(like $S=\bar{S}+c_m \chi_+$)
would restore the explicit covariance. The important detail is that the shift is
not equal for all the scalar fields of the multiplet,  
but proportional to the quark mass matrix $\mM$  
and different for each resonance.

The part of the lagrangian containing  the  vectors, axials and
pseudo-scalar resonances, $\mL_R[V,A,P]$, remains unchanged under the shift 
but   the remaining 
$\cO(p^2)$ chiral term $\mL_{2\chi}$ and 
the scalar pieces $\mL_S^{\mathrm{Kin}}\, + \, \mL_{2S}$  become 
\be
\mL_S^{\mathrm{Kin}}\, + \, \mL_{2S} \, + \, \mL_{2\chi}
\, \, = \, \,
\mL_{\bar{S}}^{\mathrm{Kin}\, '} \, + \,\mL_{2\bar{S}}^{\, '} \, + \,
\mL_{2\chi}^{\, '}
\, - \, \Frac{8 B_0^2 c_m^2}{M_S^2} \bra \mM^2 \ket \, , 
\ee
yielding a constant term proportional to $\bra\mM^2 \ket$, 
a kinetic term structure for $\bar{S}$, 
\be 
\mL_{\bar{S}}^{\mathrm{Kin}\, '}\, = \, {1\over 2} \,
\langle \nabla^\mu \left(\bar{S}+\Frac{4B_0 c_m}{M_S^2} \mM   \right) 
\nabla_\mu \left(\bar{S}+\Frac{4B_0 c_m}{M_S^2} \mM   \right) \ket 
\, \, - \, \, \Frac{1}{2}\, \bra M^2_S \bar{S}^2\rangle\, , 
\ee
an interaction lagrangian without tadpoles, 
\be
\mL_{2\bar{S}}^{\, '}  \, = 
 \, c_d \, \bra \bar{S} u_\mu u^\mu  \ket \,
+ \, c_m \, \bra \bar{S} \left(\chi_+ - 4 B_0 \mM\right) \ket \,
\ee
and a modified   $\cO(p^2)$ $\chi$PT term, 
\bel{eq.chptprime} 
\mL_{2\chi}^{\, '} \, = \, 
\Frac{F^2}{4} \, \bra \, C_\phi^{-1} \,  
\left[  u_\mu u^\mu  \, + \,  \chi_+^{'} \right]  \ket  \, , 
\ee
provided by the matrix definitions 
\bel{eq.Cphi}
\ba{rcl} 
\chi_+^{'} \, &=& \,\left( 
1 + \Frac{4 c_d c_m}{F^2} \Frac{4  B_0  \mM }{M_{S}^2} 
\right)^{-1}  \left(  1  +  
\Frac{4 c_m^2}{F^2} \Frac{4 B_0 \mM}{M_{S}^2}
\right)  \chi_+ \, ,
\\ \\ 
C_\phi^{-1}\, &=&\,  \left(1  
+ \Frac{4 c_d c_m}{F^2} \Frac{4  B_0 \mM }{M_{S}^2}\right) \, .
\ea
\ee

In order to convert the  pNGB kinetic term  to the canonical form, one needs  
to perform a re-scaling $C_\phi$ on the pNGB fields 
$u=\exp{\left(i \Phi/\sqrt{2} F \right)}
=\exp{\left(i C_\phi^\frac12 \Phi^{(can)}/\sqrt{2} F \right)}$.  

At LO in $1/N_C$ the resonance couplings are fixed 
by the QCD short distance constraints~\cite{PI:02,Jamin}: 
$c_d=c_m=F/2$. Hence, from Eqs.~\eqn{eq.chptprime} and~\eqn{eq.Cphi} 
at LO in $1/N_C$ one gets  for the pion and kaon fields the re-scalings and 
masses  
\bel{eq.canonicaLO}
 \left\{ \ba{rcl}
m_\pi^2 \, &=& \, 2 B_0 \hat{m} \, ,
\\  \\
C_\pi\, &=& \, \left[ 1  
+  \Frac{2 m_\pi^2}{M_{S}^2}\right]^{-1} \, , 
\ea\right. 
\hspace*{3cm} 
 \left\{\ba{rcl} 
m_K^2 \, &=& \,  B_0 (\hat{m} + m_s) \, ,
\\ \\
C_K\, &=& \, \left[ 1  
+  \Frac{2 m_K^2}{M_{S}^2}\right]^{-1} \, .
\ea\right. 
\ee

\subsection{$F_\pi$ and $F_K$ at Leading Order}
\tab
Since the scalar tadpole has been removed, at LO there 
is only one diagram contributing to the pion decay constant: 
the tree-level production of the pNGB from the axial current, 
\be
\bra 0| \, \bar{d} \gamma_\mu \gamma_5 u \, |P^+(p)\ket  \, \, = \, \, 
i\, \sqrt{2} \, F \, C_P^{-\frac12} \, p_\mu \, , 
\ee
and therefore the pion and kaon decay constants are 
\bel{eq.resulFP}
F_\pi\, \, = \, \, F \, \left(1+\Frac{2m_\pi^2}{M_{S}^2}\right)^\frac12 
\quad , \qquad \qquad 
F_K \, \, = \, \, F \, \left(1+\Frac{2m_K^2}{M_{S}^2}\right)^\frac12 \, . 
\ee
When  $m_P^2\ll M_S^2$ the decay constants may be  expanded 
in powers of $m_P^2$, recovering  the tree-level  $\chi$PT result 
$F_P\, =\, F \, \left[1+ \frac{4 L_5}{F^2} m_P^2+
\frac{4L_4}{F^2}(2m_K^2+m_\pi^2)+\cO(m_P^4)\right]$, with
$L_5=F^2/4M_S^2$ and $L_4=0$~\cite{PI:02}. This 
explains why the linear extrapolations work so well. 
Only near the zero quark mass the LO result in $1/N_C$ gains sizable
non-analytic contributions from the one-loop logarithms 
(NLO in $1/N_C$): 
$F_\pi\, = \, F \, 
\left[1+\frac{4 L_5^r(\mu)}{F^2} m_\pi^2
-\frac{m_\pi^2}{16\pi^2 F^2}\ln{\frac{m_\pi^2}{\mu^2}}+...\right]    $.

\begin{figure}[t!]
\begin{center}
\includegraphics[width=8.3cm]{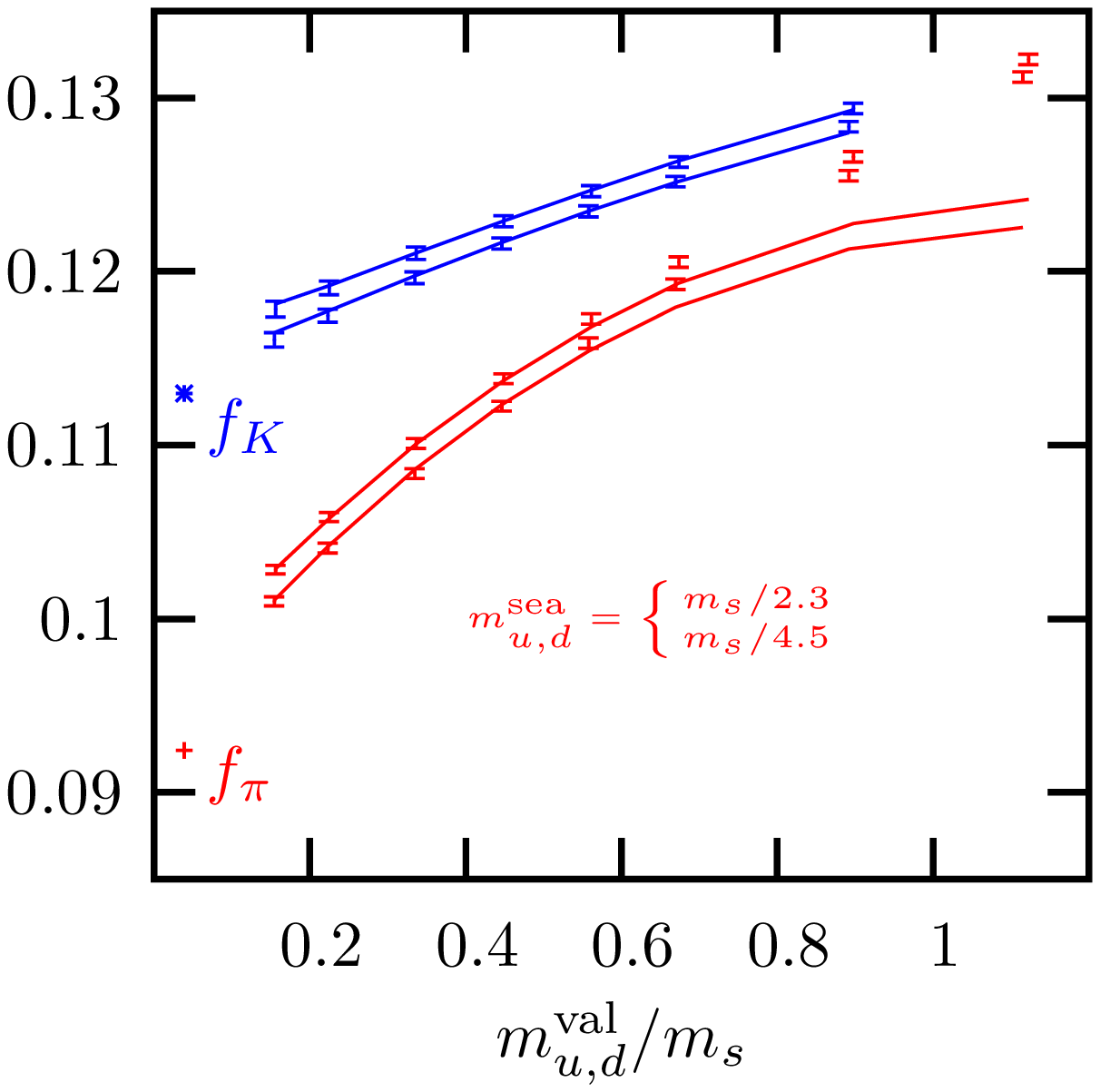}
\includegraphics[width=7.5cm,clip]{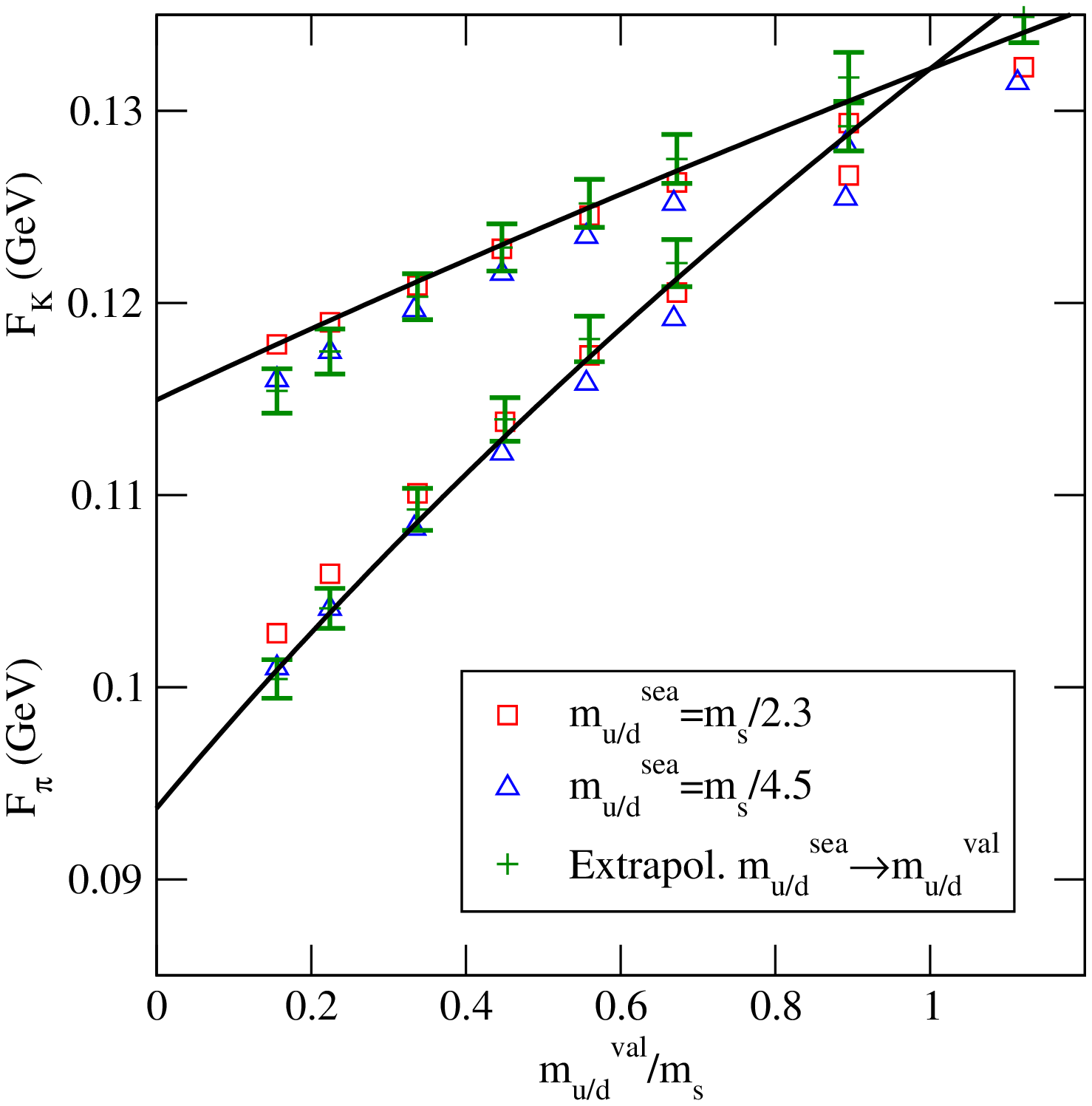}
\caption{\small{{\bf A)} Lattice results for $F_\pi$ and $F_K$ 
and  $\chi$PT extrapolations taken from MILC collaboration~\cite{latticedata}. 
{\bf B)} Comparison of the Lattice results~\cite{latticedata} with 
the R$\chi$T extrapolations for
the values from the fit, $F=93.7$ MeV, $M_S=1020$ MeV and 
the splitting $e_m^S=-0.02$.  
The kaon mass varies  as 
$m_K^2=(m_K^{Phys})^2+[m_\pi^2-(m_\pi^{Phys})^2]/2$ due to its light quark
content. They are shown together with the data for $m_{u,d}^{sea}=m_s/2.3$ 
(squares), 
$m_{u,d}^{sea}=m_s/4.5$ (triangles) and the linear extrapolation to the value 
$m_{u,d}^{sea}=m_{u,d}^{val}$ (error bars).}}
\label{fig.latticedata}
\end{center}
\end{figure}

The results for $F_\pi$ and $F_K$ from the  Lattice simulations 
(MILC Collaboration~\cite{latticedata}) are shown in   Fig.~(1.A).   
The simulation handles two kinds of quark masses: The sea-quark masses  
of the fermions within closed loops; and the valence-quark masses
of those which are not from the sea. 
In this  simulation the strange quark valence-mass 
$m_s^{val}$  and  the strange quark sea-mass $m_s^{sea}$  
take the same value 
$m_s=m_s^{val}=m_s^{sea}=1.14\;  m_s^{Phys}$, being 
$m_s^{Phys}$  the physical mass of the strange quark. It may be related 
with the physical kaon mass through 
Eq.~\eqn{eq.canonicaLO}.      
The $u/d$ quark valence-masses $m_{u,d}^{val}$ (isospin limit is assumed) 
are varied continuously between nearly zero and
the physical mass of the strange quark. Finally, the simulation 
is run for two values of the sea-masses of the quarks $u/d$: 
$m_{u,d}^{sea}=m_s/2.3$ and $m_{u,d}^{sea}=m_s/4.5$ (squares and triangles 
respectively in Fig.~(1.B)).

The modifications due to the sea quark mass are much  smaller than those from 
the valence masses, as it is expected in the large $N_C$ limit, since  
the closed quark loops would be suppressed by $1/N_C$.  
Thus,  in this work I have generated 
the matrix elements for equal values of the sea
and valence masses, i.e. for $m_{u,d}^{sea}=m_{u,d}^{val}$, through a  
simple linear extrapolation from the 
samples $m_{u,d}^{sea}=m_s/2.3$ and $m_{u,d}^{sea}=m_s/4.5$ 
(error bars in Fig.~(1.B)).

The theoretical expressions derived from R$\chi$T  for $F_\pi$ and $F_K$ 
in Eqs.~\eqn{eq.resulFP}  
were fitted to these ex\-tra\-po\-lated points. The error in my input data in the fit  
was  a  1 \% error, a typical discretization error, but it did  not  
account the uncertainties
in the  $m_{u,d}^{sea}$ extrapolation  to 
the value  $m_{u,d}^{sea}=m_{u,d}^{val}$. 
The variation of the decay constant when changing $m_{u,d}^{sea}$ could be
considered as a rough estimate of this uncertainty.  In addition, the errors due to
NLO contributions in $1/N_C$  have not been considered.    
The fit  yields the values  $F=94.1\pm 0.9$ MeV and  
$M_S=1049\pm 25$ MeV, with $\chi^2$/dof$=11.0/13$. This yields for the 
physical pion and kaon decay constants $F_\pi=95.8\pm 0.9$ MeV and 
$F_K=113\pm 1.4$ MeV, in acceptable  agreement with the experimental values  
$F_{\pi^+}=92.4\pm 0.07 \pm 0.3$ MeV 
and $F_{K^+}=113.0\pm 1.0 \pm 0.3$ MeV~\cite{PDG}. 
One may also estimate the $\chi$PT coupling at LO in $1/N_C$,  
$L_5=F^2/4M_S^2=(2.01\pm 0.10)\cdot 10^{-3}$.

Nevertheless, here the chiral logarithms are lacking. In $\chi$PT they produce 
an important non-analytic effect and a large bending in the  $F_\pi$ curve, 
since its  slope becomes large at small pion mass 
due to $\frac{dF_\pi}{dm_\pi^2}\sim \ln{\frac{m_\pi^2}{\mu}}$.  
The  effect of the logs on $F_K$ is much more reduced since 
the value of the
kaon mass does not become small when $m_{u,d}\to 0$. That is the reason
for the  better agreement of the $F_K$ result. Eventually the NLO
calculation in $1/N_C$ (one loop) would introduce 
this extra non-analytic curvature and  
the usual one loop result for  $F_\pi$ in $\chi$PT would be recovered.

The mass splitting between the masses of the two 
$I=0$ scalar resonances can also be studied since they do not contribute equally
to $F_\pi$ and $F_K$. 
The discussion is developed in the appendix. At LO in $1/N_C$ there would be 
two mass eigenstates with square masses 
$\bar{M}_{S_n}^2\equiv M_S^2- 8 e^S_m B_0 m_{u,d}$ 
and 
$\bar{M}_{S_s}^2\equiv M_S^2-8 e^S_m B_0 m_s$, with quark contents  
$\left(\frac{1}{\sqrt{2}}\bar{u}u+\frac{1}{\sqrt{2}}\bar{d}d\right)$ 
and $\bar{s}s$ respectively.   
The pion and kaon decay 
constants depend now on the splitting parameter $e^S_m$. However the 
fit to the former data is not sensitive to this coupling, yielding: 
$F=93.7\pm 1.5$ MeV, $M_S=1020\pm 80$ MeV and $e^S_m=-0.02\pm 0.05$, 
with similar $\chi^2$/dof$=10.9/12$. It provides an estimate  of 
the scalar masses at large $N_C$:  one gets the  values 
$\bar{M}_{S_n}=1020\pm 80$ MeV 
and   $\bar{M}_{S_s}= 1040\pm 90$ MeV, highly correlated, and their 
splitting, equal to $\bar{M}_{S_s}-\bar{M}_{S_n}= 20\pm 40$ MeV.  
In addition, large OZI-rule violations could occur at
NLO of the same size as the quark mass corrections. In the worst case, one 
would expect them to be of the order of $1/N_C\simeq 33$\% times 
the value of the scalar mass at LO in $1/N_C$.

The R$\chi$T extrapolation is shown in Fig.~(1.B) (solid line).  
Since this effective field theory 
is for equal valence and sea masses,  the theoretical result  
is  compared with   
an emulation of the Lattice data for  $m_{u,d}^{sea}=m_{u,d}^{val}$, 
obtained by linear extrapolation as it was explained before.

\section{Conclusions}
\tab
R$\chi$T has been shown to be an interesting tool to analyse the Lattice 
data, which usually are generated for non-physical values of the 
light quark masses. The present  study  hints that the light mesonic 
resonances may play an important
role in the large quark mass extrapolations. 
This work explores  the pion and kaon 
decay constants, providing successful results. 
Its  importance and aim  
is not just the determination of  the decay constants   
but to present an alternative  idea  to interpret  the   
Lattice simulations  for large  unphysical values of the masses,   
giving a clear explanation of      
why the usual linear extrapolations  yield such a good result. 
Likewise, this provides solid theoretical foundations based on the underlying
QCD for these techniques. 
Thus,  the  $1/N_C$ 
expansion might be as well a suitable framework  to  
describe the heavy quark matrix elements ($f_B,\, B_B...$) 
at large values of the $u/d$  quark masses, where a similar linear 
behaviour  has been also observed.

The fact that  at low energies R$\chi$T recovers $\chi$PT 
ensures that we are introducing the proper low mass 
behaviour~\cite{the role,spin1fields,quantumloops}. 
Former works~\cite{fpiextra,donoghue}  noticed 
the necessity of a separation scale $\Lambda_{_{cut}}$ where the $\chi$PT loops 
become  irrelevant.   The resonance masses provide a 
``\,natural''\, scale where the chiral extrapolations fails  
and where the dynamics of the observable changes drastically.

The fits to the simulations were done for 
an emulation of the Lattice data, 
obtained by extrapolating $m_{u,d}^{sea}$ to  
the value  $m_{u,d}^{sea}=m_{u,d}^{val}$, that    
varied in a wide and continuous range between zero and the strange quark mass. 
For  a more proper analysis one would need a simulation with equal  
sea and valence masses. 
However,   the main dependence comes from the valence-quarks and 
the sea-quark effects are small, since the closed quark loops  
are suppressed by $1/N_C$. Therefore, 
the present  calculation can be considered    
an adequate  estimate of the hadronic parameters.  
From the former fit
the values  $F=94.1\pm 0.9$ MeV and $M_S=1049\pm 25$ MeV were obtained, together
with  the $\chi$PT coupling estimate at LO in $1/N_C$,  
$L_5=(2.01\pm 0.10)\cdot 10^{-3}$. The scalar mass splitting showed  
large uncertainties.

\section*{Acknowledgments}
\tab 
I would like to thank the useful comments of A. Pich, I. Rosell and J.
Portol\'es.  Also thanks to J. F. Donoghue and P. Lepage.  
This work has been supported in part by the EU HPRN-CT-2002-00311 (EURIDICE),  
by MCYT (Spain) under grant FPA2001-3031 and by ERDF
funds from the European Commission.

{\large\bf Appendix: Scalar mass splitting}

The resonance multiplet has been considered at first 
as degenerate in mass. Nonetheless the resonance masses can gain 
contributions due to the quark masses. In the large $N_C$ limit, 
Chiral invariance requires that at order $\cO(m_q)$ 
the mass splitting  comes only through a chiral invariant term~\cite{MS}: 
\be
\Delta \mL_{m_q} \, = \,  e^S_m \bra \chi_+ S^2 \ket \, ,  
\ee
being $e^S_m$ an $\cO(N_C^0)$ dimensionless constant,  
independent of the quark masses.  

The shift now  in the scalar field to re-absorb the tadpole is slightly
different:   
\be 
S\, =\, \bar{S}\,+  \,4 B_0 c_m \mM\, \left[M^2-8 e^S_m B_0 \mM\right]^{-1} 
\, . 
\ee

The mass eigenvalues for the
$I=0$ scalars are not $M_S^2$ anymore but the two values $\bar{M}_{S_n}\equiv 
M_S^2 -8 e^S_m
B_0 m_{u,d}$, for the
state $\left(\frac{1}{\sqrt{2}}\bar{u}u+\frac{1}{\sqrt{2}}\bar{d}d\right)$, and 
$\bar{M}_{S_s}\equiv M_S^2-8 e^S_m B_0 m_s$, for $\bar{s}s$. Nonetheless, 
the physical scalar states will 
separate away  from this ideal mixing and these masses will gain 
also contributions   due to  NLO effects in $1/N_C$.

At LO in $1/N_C$ the pNGB masses still remain as given in
Eq.~\eqn{eq.canonicaLO}. However 
the re-scaling factors for pions and kaons change accordingly:
\be
C_\pi=\left[ 1+ \Frac{2 m_\pi^2}{M_S^2-4 e^S_m m_\pi^2}\right]^{-1} 
\quad , \qquad \qquad 
C_K=\left[ 1+ \Frac{2 m_K^2}{M_S^2 -4 e^S_m  m_K^2}\right]^{-1} \quad , 
\ee
which therefore modify the pion and kaon decay constants.

\end{document}